\documentstyle[prb,aps,psfig,floats]{revtex}
\begin{document}
\draft
%
\twocolumn[
\title{\vspace{-40pt}{\normalsize\null\hspace{5in}\sf 
 {\it Phys.\ Rev.\ B} (01Jun99)}\vspace{15pt}\\ 
Glassy nature of stripe ordering
in La$_{1.6-x}$Nd$_{0.4}$Sr$_x$CuO$_4$}
\author{J. M. Tranquada}
\address{Physics Department, Brookhaven National Laboratory, Upton, NY
11973-5000}
\author{N. Ichikawa and S. Uchida}
\address{Department of Applied Physics, The University of Tokyo,
Yayoi 2-11-16, Bunkyo-ku,\\ Tokyo 113, Japan}
\date{October 16, 1998}
\maketitle
\widetext
\advance\leftskip by 57pt
\advance\rightskip by 57pt

\begin{abstract}

We present the results of neutron-scattering studies on various aspects of
crystalline and magnetic structure in single crystals of
La$_{1.6-x}$Nd$_{0.4}$Sr$_x$CuO$_4$ with $x=0.12$ and 0.15.  In particular, we
have reexamined the degree of stripe order in an $x=0.12$ sample. 
Measurements of the width for an elastic magnetic peak show that it
saturates at a finite value below 30~K, corresponding to a
spin-spin correlation length of 200~\AA.  A model calculation indicates that
the differing widths of magnetic and (previously reported) charge-order peaks,
together with the lack of commensurability, can be consistently explained by
disorder in the stripe spacing.  Above 30~K (i.e., above the point at which a
recent muon spin-rotation study has found a loss of static magnetic order), the
width of the nominally elastic signal begins to increase.  Interpreting the
signal as critical scattering from slowly fluctuating spins, the temperature
dependence of the width is consistent with renormalized classical behavior of a
2-dimensional anisotropic Heisenberg antiferromagnet.  Inelastic scattering
measurements show that incommensurate spin excitations survive at and above
50~K, where the elastic signal is neglible.  Given that the stripe order is
believed to be pinned by the low-temperature tetragonal (LTT) crystal
structure, we have also investigated the transition near 70~K from the
low-temperature orthorhombic (LTO) structure.  We show that our
$x=0.12$ crystal passes through an intervening less-orthorhombic phase, before
reaching the LTT at $\sim40$~K, whereas the $x=0.15$ crystal goes directly
from LTO to LTT, with coexistence of the two phases over a range of
$\sim7$~K.  Sharp Bragg peaks in the LTT phase of the $x=0.15$ crystal
indicate a domain size of $\gtrsim1000$~\AA, with no obvious evidence for LTO
domains; hence, the coexistence of stripe order and superconductivity in this
sample cannot be explained by a mixture of crystalline phases.  Finally, we
present scattering evidence for small LTT-like domains in the LTO phase of the
$x=0.15$ sample.  A correlation between the volume fraction of such domains
and deviations of in-plane resistivity from linear $T$ dependence suggest that
charge stripes interact with these domains within the LTO matrix.

\end{abstract}
\pacs{75.50.Ee, 75.30.Fv, 71.45.Lr, 71.27+a} 
]

\narrowtext

\section{ Introduction }

Important insights into the spatial correlations of charge and spin in cuprate
superconductors have been provided by studies of the $x=\frac18$ anomaly
originally discovered\cite{mood88} in La$_{2-x}$Ba$_x$CuO$_4$.  Early on it
was demonstrated by X-ray diffraction experiments that the anomalous
suppression of superconductivity near $x=\frac18$ is associated with a
modification of the crystal structure from the usual low-temperature
orthorhombic (LTO) phase to the low-temperature tetragonal (LTT)
variant.\cite{axe89,suzu89}  After it was found that the same structural
modification can be induced in La$_{2-x}$Sr$_x$CuO$_4$ by partial substitution
of Nd for La,\cite{craw91} intensive efforts were focused on the latter
system.  A systematic study of the phase diagram revealed that not only the
symmetry of the lattice modulation but also its amplitude is correlated
with suppression of superconductivity and the appearance of local, static
magnetism.\cite{buch94a}  The discovery by neutron
diffraction\cite{tran95a,tran96b} of elastic superlattice peaks corresponding to
two-dimensional charge and spin order in a sample of
La$_{1.48}$Nd$_{0.4}$Sr$_{0.12}$CuO$_4$ suggested a likely explanation for the
anomaly: the dopant-induced charge carriers, whose density is spatially
modulated in a periodic fashion similar to an array of stripes, can be pinned
by the lattice modulation in the LTT phase, with the pinning being strongest
when the periodicity is commensurate, near $x=\frac18$.  With the charge
stripes localized, the Cu moments in intervening regions can order
antiferromagnetically, but with neighboring domains having an antiphase
relationship induced by the charge.  The suppression of superconductivity by
the stripe pinning is compatible with at least one theoretical model for the
cuprates,\cite{emer97} in which the development of superconducting phase
coherence is limited by the ability of hole pairs to tunnel between stripes. 
The Josephson coupling between stripes should be depressed by pinning.

There is a clear connection between the static correlations found in
La$_{1.6-x}$Nd$_{0.4}$Sr$_x$CuO$_4$ and the dynamic ones observed in good
superconductors.  For example, the magnetic fluctuations in superconducting
La$_{2-x}$Sr$_x$CuO$_4$ are characterized by a doping-dependent incommensurate
wave vector,\cite{cheo91} and, for a given Sr concentration, the wave vectors
found in samples with and without Nd are essentially
identical.\cite{yama98,tran97a}  Also, elastic incommensurate magnetic peaks
have now been observed in other superconducting systems, such as
La$_{2-x}$Sr$_x$Cu$_{1-y}$Zn$_y$O$_4$ (Ref.~\onlinecite{suzu98,hiro98,kimu98})
and La$_2$CuO$_{4+\delta}$ (Ref.~\onlinecite{lee98}). Furthermore, it has
recently been shown that the magnetic fluctuations in underdoped
YBa$_2$Cu$_3$O$_{6+x}$ have an incommensurate component,\cite{dai98,mook98} and
that the spatial orientation of the modulation wave vector is identical with
that in La$_{2-x}$Sr$_x$CuO$_4$.  Thus, it may be possible to gain a better
understanding of the dynamic correlations in the good superconductors by
studying the static correlations in the Nd-doped system with depressed
superconductivity.

A number of intriguing features have already been observed in the
stripe-ordered phases.  Contrary to observations on 
Fermi-surface-driven charge-density-wave (CDW) ordered systems,\cite{grun88}
infrared reflectivity studies indicate that there is no significant gap in the
optical conductivity within the CuO$_2$ planes.\cite{taji98}  Also, there is
increasing evidence that bulk superconductivity and static stripe order can
coexist.\cite{tran97a,oste97,mood97,nach98}  Given these surprising results, it
is important to investigate further the nature of the static stripe
correlations.  What is the nature of the order, both of the spins and of the
underlying lattice modulation?

We have previously shown that the spin and charge order in
La$_{1.6-x}$Nd$_{0.4}$Sr$_x$CuO$_4$ with $x=0.12$ is
quasi-two-dimensional.\cite{tran96b,vonz98}  Here we show that the ordering is
somewhat glassy.  Using high $Q$ resolution measurements of an elastic
magnetic peak, we find that the peak width $\kappa$ saturates below
$\sim30$~K, at a value corresponding to a spin-spin correlation length
($=1/\kappa$) of $\sim200$~\AA.  A model calculation, presented in the
discussion section, demonstrates that the differing spin and charge-order peak
widths\cite{vonz98}, together with the deviation from commensurability, are
consistent with disorder in the charge-stripe spacing.  The cause of the
disorder is undetermined, but might be due to interactions between the holes
and the Sr$^{2+}$ dopant potentials.

A recent muon-spin-rotation ($\mu$SR) study\cite{nach98} indicates that, for
$x=0.12$, the static magnetic correlations disappear at approximately 30~K.  In
contrast, the nominally elastic magnetic peak detected with neutrons does not
disappear until $\sim50$~K; however, we find that the peak width begins to
grow for $T\gtrsim30$~K.  The $\mu$SR result implies that the higher
temperature neutron signal involves an integration over low frequency spin
fluctuations.  Under the assumption that the finite energy resolution
integrates over the dominant critical fluctuations, $\kappa$ corresponds to
the inverse of the instantaneous spin-spin correlation length.  The
temperature dependence of $\kappa$ is consistent with the
renormalized-classical regime of an anisotropic Heisenberg
antiferromagnet.\cite{neto96,vand98}  A fit to the data yields an estimate of
the anisotropy of the effective magnetic exchange; however, in order to obtain
consistency with the narrow $Q$-width observed for spin fluctuations at
$\hbar\omega=3$~meV, it is necessary to consider quantum effects such as those
predicted when the magnetic domains correspond to even-legged spin
ladders.\cite{twor99,kim99}

We have also made a detailed study of the LTO to LTT transition in crystals
with $x=0.12$ and 0.15.  One significant result is that, for the $x=0.15$
sample, the LTT phase exhibits sharp Bragg peaks, with no obvious evidence for
LTO domains.  It follows that the stripe order and superconductivity
observed\cite{tran97a,oste97,nach98} in this sample must come from the same
crystallographic phase.  In contrast, we do observed diffuse scattering
consistent with LTT-like domains within the LTO phase for $T\gtrsim80$~K.  The
volume fraction of such domains is correlated with a deviation of the in-plane
resistivity from linear $T$ dependence.  This correlation suggests that charge
stripes may interact with the LTT-like domains, thus affecting the resistivity.

The rest of the paper is organized as follows.
After a brief description of experimental details in the next
section, characterizations of the LTO-to-LTT structural transition are presented
in section III.   In section IV, the high $Q$ resolution study of elastic
magnetic scattering for $x=0.12$ is reported.   
Some inelastic magnetic scattering measurements for the x=0.12 and 0.15
compositions are presented in section V.  In section VI, elastic diffuse 
scattering measurements near a forbidden superlattice position in the LTO phase
of the $x=0.15$ sample are analyzed.  Further discussion of the results,
including the stripe-disorder model calculation, is presented in section VII.

\section{Experimental Details and Notation}

The crystals of La$_{1.6-x}$Nd$_{0.4}$Sr$_x$CuO$_4$ used in this work were grown
at the University of Tokyo by the traveling-solvent floating-zone method.  All
are cylindrical rods 4~mm in diameter, with varying lengths.  The elastic
scattering studies of the $x=0.12$ composition were performed on crystal U2,
which is approximately 10~mm in length.  The corresponding inelastic
measurements were done on crystal U3, which was originally 40~mm in length, but
which eventually fell into pieces due to hydration of small, poorly reacted
inclusions.  Pieces of the latter crystal were used in the X-ray diffraction
study of charge-order scattering reported in Ref.~\onlinecite{vonz98}.  For the
$x=0.15$ composition, crystal U6 (35~mm in length) was used.  An earlier
crystal of the same composition (U4) was used for the measurements reported in
Ref.~\onlinecite{tran97a}.  That crystal suffered the same fate as U3, and a
piece of it was later used for the magnetization study described in
Ref.~\onlinecite{oste97}.

The neutron-scattering studies were performed on triple-axis spectrometers H4M
and H7 at the High Flux Beam Reactor, located at Brookhaven National
Laboratory.  Pyrolytic graphite (PG) monochromators and analyzers were used,
and the (002) reflection was employed in all cases, except for certain elastic,
high $Q$-resolution measurements of strong Bragg peaks, where the analyzer was
set to the (004) reflection.  Essentially all measurements were done with an
incident neutron energy of 14.7~meV, with one or more PG filters in the
incident beam to minimize the neutron flux at higher-harmonic wavelengths. 
Each crystal was cooled in a Displex closed-cycle He refrigerator, with the
temperature monitored by a Si diode.

As mentioned in the introduction, the three crystal structures of relevance to
the present work are the LTO, LTLO, and LTT, which correspond to the $Bmab$,
$Pccn$, and $P4_2/ncm$ space groups, respectively.\cite{craw91,axe94}  All of
these structures have a unit cell with lattice vectors {\bf a} and {\bf b}
rotated by
$45^\circ$ with respect to the Cu-O bonds in the CuO$_2$ planes.  When we
specify a reflection in reciprocal lattice units $(2\pi/a_o,2\pi/b_o,2\pi/c)$
with respect to this cell, we will denote it with a subscript $o$ for
``orthorhombic''.  (In the LTT phase at 10~K, $a_o=5.34$~\AA and $c=13.1$~\AA.)
To describe the magnetic scattering, it is more convenient to work with respect
to a smaller cell with lattice vectors parallel to the in-plane Cu-O bonds.  In
this case, the reciprocal lattice units are
$(2\sqrt{2}\pi/a_o,2\sqrt{2}\pi/b_o,2\pi/c)$, and no subscript will be
appended to the reciprocal-space coordinates.

\section{LTO-to-LTT Transition}

\begin{figure}
\centerline{\psfig{figure=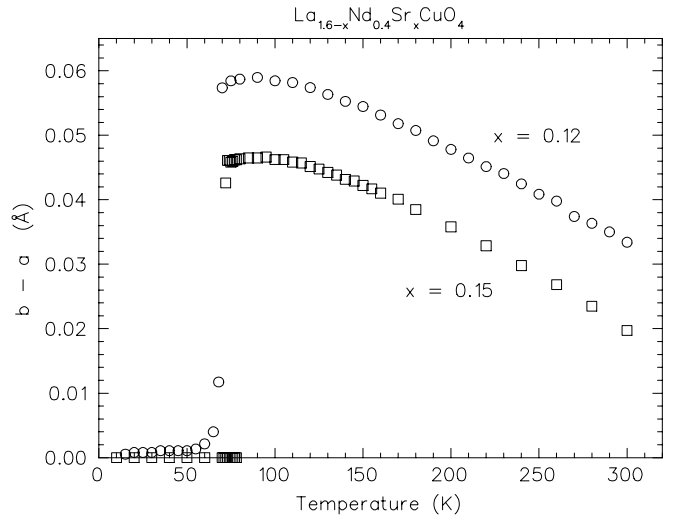,width=3.4in}\medskip}
\caption{Temperature dependence of the difference in in-plane lattice
parameters, $b-a$ (based on LTO unit cell), for the
La$_{1.6-x}$Nd$_{0.4}$Sr$_x$CuO$_4$ crystals with $x=0.12$ (circles) and
$x=0.15$ (squares).}
\label{fg:b_a}
\end{figure}

In order to interpret the nature of the magnetic order, it is first necessary
to characterize the underlying crystalline order.  As a starting point,
Fig.~\ref{fg:b_a} shows the difference between the in-plane lattice
parameters, $b-a$, as a function of temperature for two samples.  The $x=0.15$
crystal transforms directly from LTO to LTT, with a coexistence of 
significant fractions of the two phases between 71 and 77~K (see
Fig.~\ref{fg:400_peaks} below).  In contrast, the $x=0.12$ sample shows a
large but incomplete decrease in the orthorhombic splitting at approximately
69~K.

The measurements on the $x=0.12$ sample require a bit of explanation.  In the
LTO phase, there are four possible orientations of twin domains.  With the
large mosaic crystals commonly studied, one typically finds the simultaneous
presence of all four domain orientations, with comparable populations of the
four domains.  As a result, a scan along the $[100]_o$ direction through a Bragg
peak at $(H00)_o$ will also detect a twin-related peak corresponding to
$(0H0)_o$.  Such is the case for the $x=0.15$ sample, and it allows one to
measure the orthorhombic splitting with a single scan.  With the $x=0.12$
crystal (U2), on the other hand, a single domain dominates (and there are a
total of 3, rather than 4, domains, with unusual relative orientations).  As a
result, it is necessary to separately scan the $(H00)_o$ and $(0H0)_o$
reflections.  An advantage of this situation is that, because the peaks do not
overlap, we can achieve a better resolution of small changes in
orthorhombicity.  On the other hand, our absolute measure of orthorhombicity
is limited by the precision with which the crystal is aligned.

\begin{figure}
\centerline{\psfig{figure=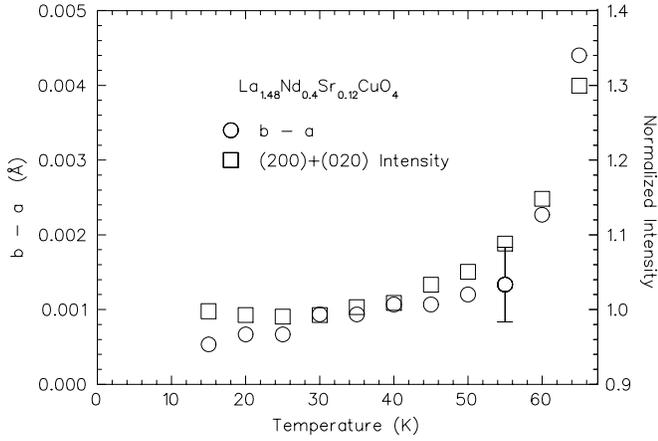,width=3.4in}\medskip}
\caption{Comparison between the orthorhombic splitting, $b-a$, (circles) and
the average intensity of the (200) and (020) Bragg peaks (squares), normalized
at low temperature.}
\label{fg:strain}
\end{figure}

The splitting measured for the $x=0.12$ crystal below the structural
transition temperature is shown in Fig.~\ref{fg:strain}.  The error bar
indicates twice the maximum variation in splitting between cooling and warming
cycles (and corresponds to 4\%\ of the peak width).  The absolute uncertainty
is certainly much greater than this, and we cannot establish from these
measurements whether the crystal is tetragonal or orthorhombic at low
temperature.  It is certain that any remaining orthorhombicity below
$\sim40$~K is extremely small, and, based on powder diffraction studies of
samples with the same composition,\cite{craw91,buch94a,mood98} we expect that
the crystal is actually tetragonal.  The significant point is that the
splitting grows monotonically between 50~K and the transition at 69~K.  This
behavior is consistent with the occurrence of an intervening LTLO phase, as
first observed by Crawford {\it et al.}\cite{craw91}.  A corresponding feature
is the increase in integrated intensity for the $(200)_o$ and $(020)_o$ Bragg
peaks (see Fig.~\ref{fg:strain}), which was seen previously in a single-crystal
study by Shamoto {\it et al.}\cite{sham92}  The most likely explanation for the
intensity variation is a reduction in extinction due to strain associated with
the growing orthorhombicity.  Assuming this to be the correct interpretation,
the intensity provides a more precise measure of the structural changes, and
it suggests that the crystal is essentially tetragonal, and no longer
changing, for $T\lesssim40$~K.

Turning to the $x=0.15$ crystal, Fig.~\ref{fg:400_peaks} shows high-resolution
elastic scans through the $(040)_o$/$(400)_o$ Bragg peaks.  At 80~K, the LTO
peaks are sharp and well separated; however, close inspection reveals a weak
and broad contribution centered midway between the strong peaks.  A similar
feature has recently been observed in an X-ray powder diffraction study by
Moodenbaugh {\it et al.},\cite{mood98} and interpreted as evidence for small
LTT-like domains.  We will return to that issue in section VI.  For now, we
focus on the sharp LTT peak that begins to appear below 78~K.  As indicated in
(b) and (c), there is coexistence of the LTO and LTT phases down to
approximately 71~K.  (Moodenbaugh {\it et al.}\cite{mood98} observed a broader
coexistence region of approximately 30~K in their powder study.)  While a
single, strong LTT peak dominates at 70~K, there is also a very weak component
with a width roughly 3 times greater that lies under it.

The point to be emphasized here is the narrow width of the dominant LTT peak
at 70~K.  It is only marginally broader than the LTO peaks at 80~K, and all
are close to the resolution width.  Taking the resolution into account, we
estimate a typical domain size of $\sim1000$~\AA\ or greater.  The LTT peak
observed in this crystal is considerably sharper than those typically detected
in powder samples,\cite{axe89,craw91,buch93} where the LTT peak width is
comparable to the LTO splitting.  The broad peaks observed in powders raised
concerns that the superconductivity which appears to survive in the LTT
phase\cite{oste97,mood97,nach98} might actually be associated with small LTO
domains.  In light of the present results, such a possibility seems quite
unlikely.

\begin{figure}
\centerline{\psfig{figure=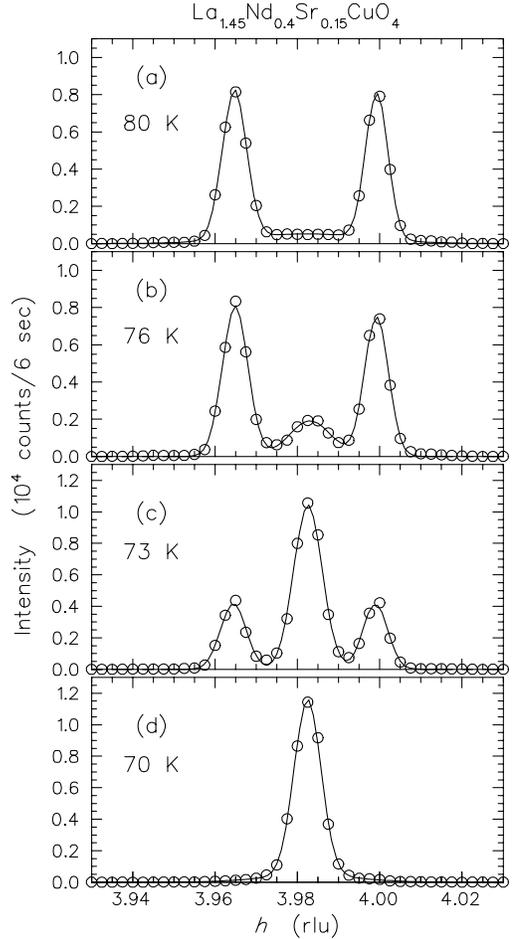,width=2.6in}\medskip}
\caption{Scans of the $(040)_o$/$(400)_o$ Bragg peaks at several
temperatures through the transition from the LTO phase (a), to the LTT phase
(d), with coexistence of the two phases at intermediate temperatures, (b) and
(c). Measured at H7 using $E_i=14.7$~meV, horizontal collimations $=$
$10'$-$10'$-$10'$-$10'$, monochromator $=$ PG(002), analyzer $=$ PG(004).}
\label{fg:400_peaks}
\end{figure}

\section{Elastic Magnetic Scattering}

As discussed elsewhere,\cite{tran95a,tran96b} magnetic superlattice peaks
are observed within the $(h,k,0)$ zone of reciprocal space below
approximately 50~K for $x=0.12$, and 45~K for $x=0.15$.  Below $\sim3$~K,
where the Nd moments start to order via coupling to the ordered Cu moments,
thus enhancing the superlattice intensity with their much larger moments, it
was possible to investigate the correlation lengths.  Within the planes, the
correlation length was found to be substantial but finite, while the magnetic 
correlations along the $c$ axis are very short-range.\cite{tran96b}  It was
difficult to investigate the in-plane correlation length at higher
temperatures because the weakness of the intensity limited the degree to
which it was practical to obtain sufficient $Q$ resolution with tightened
collimation.

\begin{figure}
\centerline{\psfig{figure=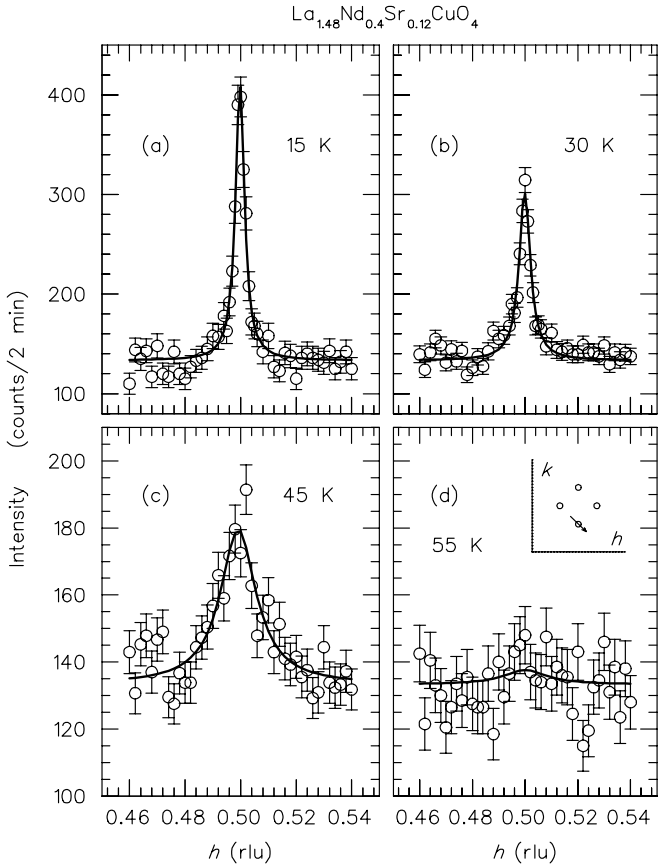,width=3.4in}\medskip}
\caption{Elastic scans in the transverse direction (see inset) through the
magnetic superlattice peak at (0.5,0.382) in
La$_{1.48}$Nd$_{0.4}$Sr$_{0.12}$CuO$_4$ for several different temperatures.  The
solid lines through the data points are fits to a Lorentzian plus constant
background.  Note the change in the intensity scale between (b) and (c). 
Measured at H4M using $E_i=14.7$~meV, horizontal collimations $=$
$40'$-$80'$-$20'$-$80'$.}
\label{fg:mag_peaks}
\end{figure}

Making another attempt, we have now succeeded in following the temperature
dependence of the peak widths by performing transverse scans, where we can
take advantage of the intrinsically tight resolution at small $Q$.  Examples
of such scans are presented in Fig.~\ref{fg:mag_peaks}.  The fitted peak shape
corresponds to a Lorentzian plus a constant background.  (No systematic
temperature dependence was found for the background, so the background value
averaged over essentially all temperatures was applied and held fixed in the
final fits.)  An alternative peak shape has been suggested based on the
recently proposed analogy between charge stripes and liquid
crystals.\cite{kive98}  If the pinned stripes behave like a smectic liquid
crystal, then one might expect the spatial correlations to decay
algebraically, rather than exponentially.\cite{chai95}  In that case, the
peak shape in reciprocal space would be proportional to $(Q-Q_0)^{-p}$. 
However, if the exponent $p$ is close to 2, one would need a much better
signal-to-noise ratio and excellent knowledge of the background in order to
distinguish the algebraic line shape from a Lorentzian.  Further experimental
advances will be required to test for such differences.

\begin{figure}
\centerline{\psfig{figure=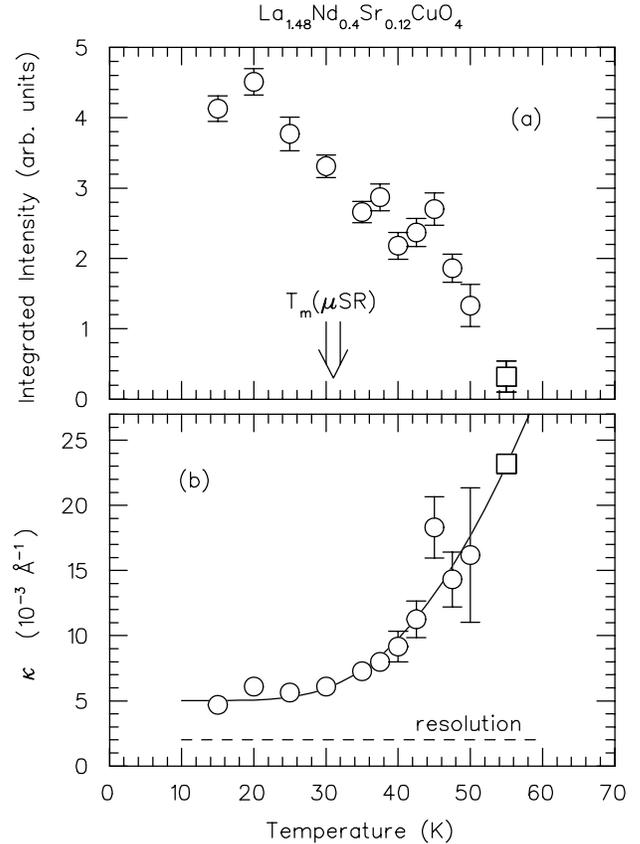,width=3.2in}\medskip}
\caption{Results of the Lorentzian fits to the magnetic peak scans, as a
function of temperature. (a) Integrated intensity (in arbitrary units); arrow
indicates the magnetic ordering temperature detected by $\mu$SR
(Ref.~\protect\onlinecite{nach98}). (b)  Peak half-width, $\kappa$; dashed
line indicates resolution half-width; line through points is discussed in the
text.  In fitting the scan at $T=55$~K (squares), the width was constrained to
lie on the curve, and only the amplitude was varied. }
\label{fg:qwid}
\end{figure}

Results for the integrated intensity and peak half-width at half-maximum,
$\kappa$, are shown as a function of temperature in Fig.~\ref{fg:qwid}.  As
indicated in (b), the resolution width is much less than the measured width,
thus justifying the absence of a resolution correction in the fitting. 
Considering first the intensity, it is interesting to compare with a recent
muon-spin-rotation ($\mu$SR) study of magnetic order in a similar
crystal.\cite{nach98}  There, a loss of order was detected near 30~K, as
indicated by the arrow in Fig.~\ref{fg:qwid}(a).  In contrast, the ``elastic''
signal seen by neutron scattering appears to survive up to approximately 50~K. 
Such an apparent conflict has been found previously\cite{ster90,keim92b} in
studies of magnetic correlations in La$_{2-x}$Sr$_x$CuO$_4$ with $x\sim0.05$. 
It can be understood in terms of the distinct sensitivities of the two
techniques to fluctuations.  In
$\mu$SR, the local hyperfine field is considered static when it does not vary
significantly on the time scale of the muon lifetime, $\sim2$~$\mu$s.  On the
other hand, the neutron measurement integrates over fluctuations within a
Gaussian resolution window with a full width at half maximum of approximately
0.8~meV (0.2~THz).  Thus, the observations indicate that between 30~K and 50~K
there is no static order, but instead there is a characteristic fluctuation
rate that is gradually increasing from $10^6$ to $10^{11}$ Hz.

The temperature dependence of $\kappa$ is qualitatively consistent with the
above scenario.  Below 30~K, $\kappa$ remains constant.  Taking the inverse of
$\kappa$ at low temperature gives a correlation length of 200~\AA.  Above
30~K, $\kappa$ starts to increase.  The curve through the data points
corresponds to a fit with the form
\begin{equation}
  \kappa = \kappa_0 + Ae^{-B/k_BT},
  \label{eq:kap}
\end{equation}
which is essentially the formula used by Keimer {\it et al.}\cite{keim92b} to
describe the inverse correlation length in lightly doped La$_2$CuO$_4$.  The
fit yields the following parameter values: $\kappa_0=0.0050(2)$~\AA$^{-1}$,
$A=0.6(6)$~\AA$^{-1}$, and $B=17(4)$~meV, where the estimated uncertainty in the
last digit is listed in parentheses.

The second term on the right-hand side of Eq.~(\ref{eq:kap}) has the form
predicted for the 2D quantum antiferromagnetic Heisenberg model in the
renormalized classical regime.\cite{chak88,chak89}  Fits to experimental
measurements of $\kappa(T)$ in pure and lightly-doped La$_2$CuO$_4$ by Keimer
{\it et al.}\cite{keim92b} yielded $B\approx J$, where $J$ is the exchange
energy between nearest-neighbor Cu spins.  It is not immediately clear that
such a formula, based on a spin-only model, should be relevant to the present
case, where there is a substantial concentration of holes in the planes. 
Castro Neto and Hone\cite{neto96} have argued that within the stripe model,
where the holes are restricted to moving along antiferromagnetic domain
walls, the main effect of the holes would be to weaken the effective exchange
between spins on either side of a domain wall.  Assuming that this weakened
coupling is the dominant factor for determining long-wavelength excitations
along the modulation direction, they obtain an effective anisotropic
Heisenberg model.  Analysis of $\kappa(T)$ based on that model yields the same
exponential dependence on the inverse temperature,\cite{neto96,vand98} as
obtained in the isotropic case, but with modifications to the relationship
between $B$ and $J$.    

Let $\alpha$ represent the value of the effective exchange perpendicular to
the stripes relative to the exchange parallel to the stripes.  Then, according
to Ref.~\onlinecite{neto96},
\begin{equation}
  B = 2\pi\rho_s(\alpha),
\end{equation}
with
\begin{equation}
  \rho_s(\alpha) = \sqrt{\alpha}JS^2,
\end{equation}
where $S=\frac12$ is the spin.  If we assume $J$ to have the same value as in
La$_2$CuO$_4$ ($\approx135$~meV\cite{hayd91}), then from the fitted value of
$B$ we find $\alpha\approx0.01$.  Going further, the theoretical expression
for $A$ is\cite{neto96}
\begin{equation}
  A\approx 8\sqrt{2}\pi\alpha^{\frac14}/ae,
\end{equation}
where $a$ is the lattice parameter.  Plugging in $\alpha\approx0.01$ gives
$A\approx0.5$, in good agreement with the fitted value.

Within the anisotropic Heisenberg model,\cite{neto96,vand98} the spin-wave
velocity perpendicular to the stripes (along the modulation direction) is
\begin{equation}
  c_{\bot}\approx\sqrt{\frac{\alpha}2}c_0,
\end{equation}
where $c_0$ is the spin-wave velocity of the undoped antiferromagnet.  The
fitted value of $\alpha$, together with the experimental\cite{aepp89} $c_0$
for La$_2$CuO$_4$, gives $\hbar c_{\bot} \approx50$~meV-\AA.  We can use this
result to check the assumption that the neutron measurement is integrating
over low-energy fluctuations for $T\gtrsim30$~K.  Neutrons measure the dynamic
susceptibility, $\chi''({\bf Q},\omega)$, multiplied by
$1/(1-e^{-\hbar\omega/k_BT})$.  Integrating over {\bf Q}, there is a peak in
this quantity at $\Gamma\approx\hbar c\kappa$.  Using the numbers above, we
find $\Gamma\approx0.2$~meV at 30~K, and $\Gamma\approx0.8$~meV at 50~K, where
the signal is disappearing.  The former value is smaller than the energy
resolution half-width, while the latter is larger.  Thus, our
interpretation appears reasonable.  On the other hand, we will show in the
next section that $\hbar c_\bot\approx50$~meV-\AA\ is much too small to be
compatible with the observed low-energy inelastic scattering.  It is also
incompatible with the range of energies over which incommensurability is
observed\cite{mats94,aepp97} in La$_{2-x}$Sr$_{x}$CuO$_4$ with $x\approx0.15$.

One solution to this dilemma is suggested by the analysis of Tworzyd\l o {\it
et al.}\cite{twor99}  They have pointed out that when the magnetic domains
become sufficiently narrow (small stripe spacing) one can view them as coupled
spin ladders.  For small $\alpha$, one obtains qualitative differences
depending on whether the spin ladders have an even or odd number of legs. 
(This distinction is absent in the original treatment of the anisotropic
Heisenberg model.\cite{neto96,vand98})  In particular, Tworzyd\l o {\it et
al.}\cite{twor99} have shown that for 2-leg ladders a quantum-disordered state
occurs for $\alpha\le0.30$.  (For 4-leg ladders,\cite{kim99} the transition is
at $\alpha=0.07$.)  In terms of the 2-leg ladder model, our observed
temperature dependence of $\kappa$ could be explained with $\alpha\approx0.35$,
which would also be consistent with relatively stiff spin waves.\cite{twor99}

Tworzyd\l o {\it et al.}\cite{twor99} have suggested that the 2-leg ladder
model could be appropriate at $x=\frac18$ if the hole stripes are centered on
O rows\cite{whit98a} in a manner such that they tie up the spins on the
immediately adjacent rows of Cu.  If this were the case, then tuning $x$ away
from $\frac18$ should introduce odd-leg ladders, which would presumably
enhance the tendency for magnetic order.\cite{kim99}  In contrast,
experiments\cite{tran97a,kuma94,kimu98} indicate that the highest magnetic
ordering temperature actually occurs at
$x\approx0.12$.  Thus, while the spin-only models can explain several features
of the measurements, questions still remain.  It is quite possible that the
fluctuations of the charge stripes, which are ignored in spin-only models,
play a significant role in determining the spin correlations.

One other puzzle concerns the saturation of $\kappa$ below 30~K.  If the
characteristic energy for critical spin fluctuations were 0.2~meV (or perhaps
higher), then most of the spin
fluctuations would be too fast to allow any muon precession to be observable. 
One possibility is that our analysis is too simple minded.  Alternatively, it
may be that the lack of temperature dependence in $\kappa$ at low $T$ is
associated with disorder in the charge-stripe array, rather than with the
temperature dependence of the spin fluctuations.  Analysis of such a disorder
model is presented in section VII.

\section{Inelastic Magnetic Scattering}

\begin{figure}
\centerline{\psfig{figure=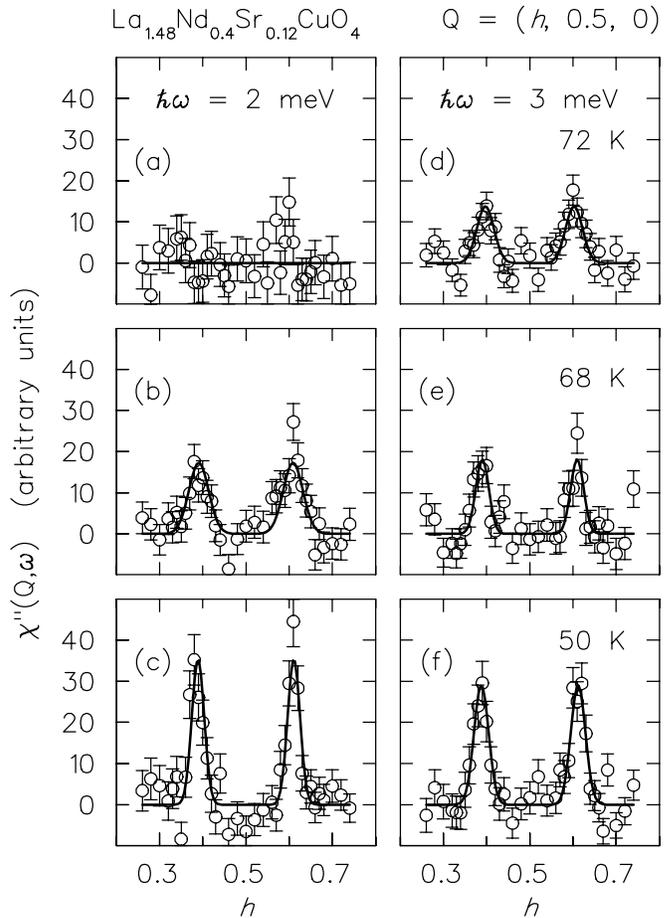,width=3.4in}\medskip}
\caption{Constant-energy scans along $Q=(h,0.5,0)$, through the magnetic wave
vectors $h=0.5\pm\epsilon$, for La$_{1.48}$Nd$_{0.4}$Sr$_{0.12}$CuO$_4$.  The
Bose factor has been divided out, a fitted background (linear in $h$) has
been subtracted, and a correction has been made for differences in resolution
volume at the two energies.  Panels on the left correspond to
$\hbar\omega=2$~meV and temperatures of (a) 72~K, (b) 68~K, (c) 50~K; on the
right, $\hbar\omega=3$~meV and temperatures are (d) 72~K, (e) 68~K, (f) 50~K.
Measured at H4M using a fixed incident energy of 14.7~meV, horizontal
collimations $=$ $40'$-$120'$-$80'$-$80'$, and
typical counting time of 20 min per point. Solid lines are gaussians fit with
the constraint that the peaks be symmetric about $h=0.5$.}
\label{fg:chi12}
\end{figure}

It is also interesting to characterize the spin dynamics directly through
measurements of inelastic scattering.  Figure~\ref{fg:chi12} shows results for
$x=0.12$ obtained by scanning through the incommensurate positions, along
$Q=(h,0.5,0)$, with the energy transfer fixed at 2~meV [(a),(b),(c)] and at
3~meV [(d),(e),(f)].  A linear background has been subtracted, corrections
have been made for differences in resolution volume at the two energies, and
the thermal factor has been divided out, so that the resulting signal is
proportional to $\chi''({\bf Q},\omega)$.  The scattering is peaked rather
sharply about the incommensurate wave vectors defined by the elastic scans,
with the $Q$-width limited by resolution at 50~K.  The signal is relatively
strong at 50~K, and gradually decreases as the temperature is raised.  At 72~K,
above the transition to the LTO phase, the signal is still finite at 3~meV. 
This last result provides an important connection with
studies\cite{mats94,aepp97} of spin fluctuations in La$_{2-x}$Sr$_x$CuO$_4$,
where the structure is LTO.  The LTT phase induced by Nd
substitution appears to stabilize a static component of correlations that
already exist dynamically in the LTO phase.

\begin{figure}
\centerline{\psfig{figure=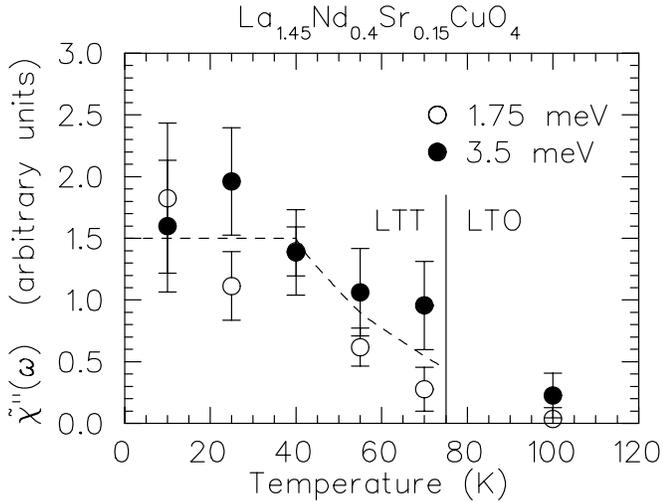,width=3.4in}\medskip}
\caption{$Q$-integrated $\chi''(Q,\omega)$ for
La$_{1.45}$Nd$_{0.4}$Sr$_{0.15}$CuO$_4$ obtained from scans similar to those in
the previous figure.  The vertical line indicates the LTT-LTO phase boundary,
and the dashed line is a guide to the eye.}
\label{fg:chi15}
\end{figure}

Inelastic measurements have also been performed on an $x=0.15$ sample. 
Results for the local susceptibility, 
$\tilde\chi''(\omega)=\int d{\bf Q}_{2D} \chi''({\bf Q},\omega)$, at
two excitation energies, 1.75 and 3.5~meV, are presented as a function of
temperature in Fig.~\ref{fg:chi15}.  $\tilde\chi''(\omega)$ changes relatively
little between 10 and 40~K, where an elastic magnetic signal is observed. 
Above the neutron-determined ordering temperature ($\sim45$~K), the signal at
1.75~meV falls off rapidly, whereas the susceptibility at 3.5~meV appears to
decrease more slowly.  This behavior is qualitatively similar to the decay of
the low-energy spin-fluctuation intensity in La$_2$CuO$_4$ above
$T_N$.\cite{yama89}  

\begin{figure}
\centerline{\psfig{figure=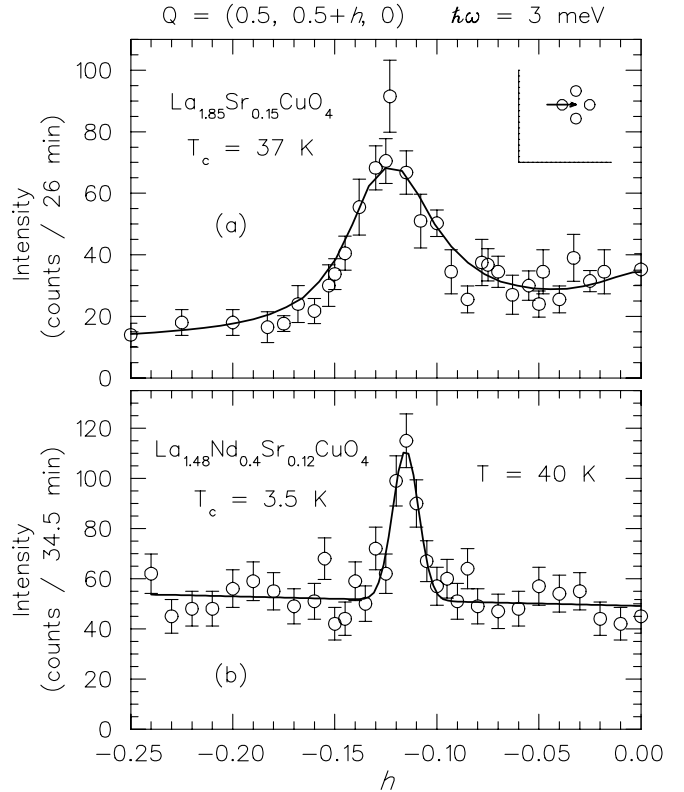,width=3.4in}\medskip}
\caption{Comparison of constant-energy scans at $\hbar\omega=3$~meV through
an incommensurate magnetic peak (see inset) for (a)
La$_{1.85}$Sr$_{0.15}$CuO$_4$ (Ref.~\protect\onlinecite{ster,yama95a}), and (b)
La$_{1.48}$Nd$_{0.4}$Sr$_{0.12}$CuO$_4$ (crystal U2).  Both measurements were
performed with a fixed incident energy of 14.7~meV and a sample temperature of
40~K.  Scan (a) was done at H7 with horizontal collimations of
$40'$-$40'$-$40'$-$80'$, while (b) was done at H4M with
$40'$-$80'$-$80'$-$80'$. Fitted curves are discussed in the text.}
\label{fg:3mev}
\end{figure}

Finally, it is interesting to compare the $Q$-width of the excitations in the
$x=0.12$ crystal with that found in La$_{2-x}$Sr$_x$CuO$_4$ with $x=0.15$. 
Figure~\ref{fg:3mev} shows scans for two such samples obtained under similar
conditions: $\hbar\omega=3$~meV and $T=40$~K.  The curve through the data in
(b) is a Gaussian plus a linear background, while the curve in (a) is a fit
based on a disordered stripe model described in Ref.~\onlinecite{tran97d}. 
The peak in (b), which is resolution limited, is clearly much narrower than
that in (a).  Such a difference appears to be consistent with the more
dynamic nature of the correlations in the $x=0.15$ sample with no Nd.  

The inelastic $Q$ width at 2~meV in La$_{2-x}$Sr$_x$CuO$_4$ with $x=0.12$ has
recently been reported by Yamada {\it et al.}\cite{yama98}  They find a value
essentially the same as the resolution-limited width that we observe at 3~meV in
the Nd-doped crystal.  An elastic magnetic signal at the same incommensurate
position has also been discovered in the $x=0.12$ crystal with no
Nd.\cite{suzu98}  Thus, there seems to be a reasonable correlation between
decreasing inelastic $Q$-widths (perhaps limited by the correlation length)
and the onset of static order.

Before moving on, we should consider whether the effective spin-wave velocity
discussed in the last section is consistent with the observed peak width
(half-width $=0.012$~\AA$^{-1}$) in Fig.~\ref{fg:3mev}(b).  If the width were
determined entirely by dispersion, rather than resolution, then the
corresponding spin-wave velocity is $\sim250$~meV-\AA.  This is almost 3 times
greater than the estimate we obtained for $\hbar c_{\rm eff}$ from
$\kappa(T)$.  Part of the discrepancy might be associated with anisotropy in
the dispersion; however, attempts to test this experimentally have so far been
limited by anisotropy in the resolution function.  Clearly, more work is
required to obtain a consistent quantitative account of the
quasi-elastic and inelastic scattering.

\section{Imperfect Structural Order in the LTO Phase}

\begin{figure}
\centerline{\psfig{figure=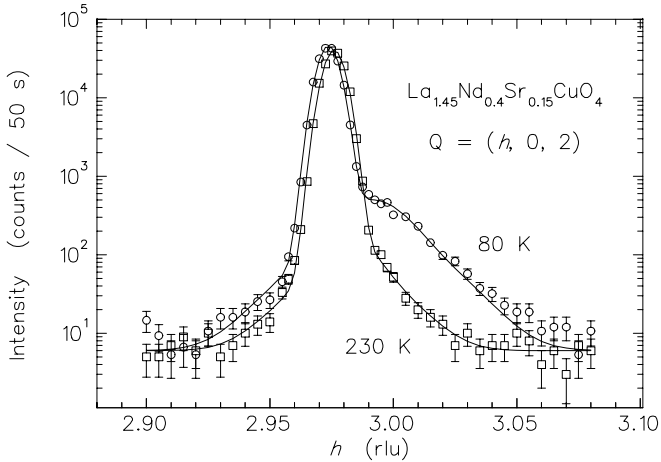,width=3.4in}\medskip}
\caption{Elastic scans along $Q=(h,0,2)$ at two different temperatures for
La$_{1.45}$Nd$_{0.4}$Sr$_{0.15}$CuO$_4$.  Note that the intensity is plotted
on a logarithmic scale.  The strong peak at $h\approx2.975$ corresponds to the
LTO-allowed (032) superlattice peak.  The broad feature centered near $h=3$ at
80~K is diffuse scattering associated with the (302) reflection, which is
forbidden in LTO but allowed in LTT and LTLO.  The fitted curves are discussed
in the text.}
\label{fg:032}
\end{figure}

We now return our attention to the structural order of the $x=0.15$ crystal in
the LTO phase.  In section III we noted the presence of a broad, unexpected
peak between the $(040)_o$ and $(400)_o$ Bragg reflections [see
Fig.~\ref{fg:400_peaks}(a)].  The position of the broad peak suggests
a tetragonal (or less-orthorhombic) component, and the width corresponds
to domains of width $\sim140$~\AA.  To test whether this component is
consistent with LTT-like domains, we looked for related scattering near a
characteristic superlattice position.  Figure~\ref{fg:032} shows scans along
${\bf Q}=(h,0,2)_o$ (note the logarithmic intensity scale).  The strong peak
at $h\approx2.975$ is the allowed $(032)_o$ of the LTO phase; the noninteger
value of $h$ is due to the slight orthorhombic splitting between $a^\ast$ and
$b^\ast$.  In the 80~K scan, one can see a second peak, broad and weak, that
is centered at $h\approx 3.00$.  Such a peak is not allowed in the LTO phase,
but would be expected in the LTT.  Its intensity has diminished considerably in
the scan at 230~K.

To analyze the temperature dependence of the diffuse $(302)_o$ scattering, we
restricted our choice of fitting functions to Gaussians.  (Lorentzians were
rejected because of the significant peak area associated with tails outside of
the measurement region.)  A single gaussian was used for the $(032)_o$ peak,
but was found to give a poor fit to $(302)_0$.  A more satisfactory fit was
obtained using the following combination of two Gaussians:
\begin{equation}
  I(h) = A\left[e^{-\frac12{(h-h_0)^2/\sigma^2}} +
         {\textstyle\frac12} e^{-\frac12{(h-h_0)^2/(2\sigma)^2}}\right].
\end{equation}
Typical fits are indicated by the curves through the data points in
Fig.~\ref{fg:032}.  At $80$~K, the peak center, $h_0$, is 2.992(2), and the
widths of the two Gaussian peaks correspond to domain sizes of 180~\AA\ and
90~\AA.  The average of these lengths is similar to the value obtained from
the broad $(400)_o$ peak.  As the peak width does not vary significantly with
temperature, the intensity reflects the volume fraction of LTT-like domains
within the LTO matrix.

\begin{figure}
\centerline{\psfig{figure=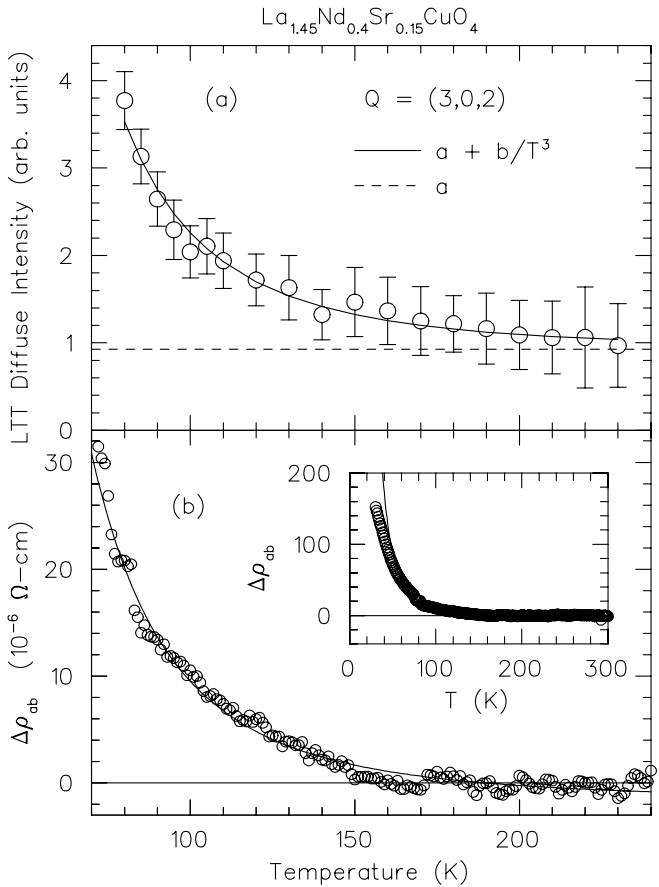,width=3.4in}\medskip}
\caption{(a) Temperature dependence of the diffuse intensity centered at
$Q=(3,0,2)$, for La$_{1.45}$Nd$_{0.4}$Sr$_{0.15}$CuO$_4$.  Solid line is a fit
to the form $I=a+b/T^3$; dashed line indicates the temperature-independent
term.  (b) In-plane resistivity remaining after subtraction of a dominant
contribution linear in $T$, determined from a fit over the range 
$170~{\rm K}\le T\le 300$~K.  The line through the data is a fit to the form
$\Delta\rho_{ab}=a+b/T^3$.  The inset shows the data and fit over a larger
temperature range.}
\label{fg:rho}
\end{figure}

The temperature dependence of the $(302)_o$ intensity is plotted in
Fig.~\ref{fg:rho}(a).  The intensity varies little at high temperature, but it
begins to grow rapidly at lower temperatures.  As indicated by the curve
through the data points, the temperature dependence is described well by a
function of the form $a+b/T^3$.  The significance of
the temperature independent term is not clear.  As one can see in the 230-K
scan in Fig.~\ref{fg:032}, the diffuse peak becomes difficult to distinguish
from possible (very weak) tails on the strong $(032)_o$ peak.  Thus, it is
possible that the scattering from LTT-like domains actually disappears at room
temperature.

It is interesting to compare the temperature dependence of the $(302)_o$
scattering with that of the in-plane resistivity.  In the LTO phase,
$\rho_{ab}$ is nearly linear in temperature, just as is observed for samples
with no Nd; however, a slight deviation from linearity begins to become
noticeable below 150~K.   To examine this deviation, we first fit the data
between 170 and 300~K to the form
\begin{equation}
  \rho_{ab} = \rho_0 +\rho_T T,
\end{equation}
finding $\rho_0=-0.01$~$\mu\Omega$-cm and
$\rho_T=1.26\times10^{-3}$~$\mu\Omega$-cm/K.  Extrapolating the linear
contribution and subtracting from the data gives $\Delta\rho_{ab}$, which is
plotted in Fig.~\ref{fg:rho}(b).  As one can see, $\Delta\rho_{ab}$ grows at
low temperature in a fashion similar to the $(302)_o$ intensity.  In fact, the
temperature dependence is well described by the same functional form, as
indicated by the curve though the data points.

Given the evident correlation between the $(302)_o$ intensity and
$\Delta\rho_{ab}$, we speculate that the rise in $\Delta\rho_{ab}$ is
associated with charge-stripe pinning in the LTT-like domains which occupy a
small volume fraction of the sample.  Electron-diffraction studies of
La$_{1.88}$Ba$_{0.12}$CuO$_4$ and La$_{1.5}$Nd$_{0.4}$Sr$_{0.1}$CuO$_4$
indicate that LTT-like domains first appear at the LTO twin
boundaries.\cite{chen93,zhu94,inou97}  If LTT-like domains uniformly decorate
all LTO twin boundaries, then it may be difficult for an electron (or hole) to
move through the sample without passing through a region of pinned or
nearly-pinned charge stripes.  The continuous development of the LTT-like
volume fraction and its impact on the resistivity may explain, at least in
part, why the resistivity does not clearly reflect the temperature dependence
of the charge order parameter determined by diffraction.\cite{tran96b,vonz98}

\section{Summary and Discussion}

We have presented neutron scattering studies of various aspects of ordering
in crystals of\linebreak
La$_{1.6-x}$Nd$_{0.4}$Sr$_x$CuO$_4$ with $x=0.12$ and
0.15.  The most significant results are the following: 1) At low temperature,
where the Cu spins order in a nearly two-dimensional stripe structure, the
magnetic correlation length remains finite.  2) The temperature dependence of
the correlation length above 30~K is similar to that of the
renormalized-classical regime of a 2D Heisenberg antiferromagnet, and applying
the formula for the spin-only system yields a measure of the effect that the
charge stripes have on the effective spin-wave velocities.  3) Within the
LTO phase, there is a correlation between the deviation of the resistivity from
a linear temperature dependence and the volume fraction of LTT-like domains. 
Below, each of these points will be discussed in turn.

\subsection{Finite correlation length}

What can we say about the nature of the magnetic disorder in the ground state of
the $x=0.12$ sample?  One possible model would be isolated, ordered domains of
finite size within a sea of disorder; however, such a picture is in conflict
with the zero-field $\mu$SR observation that essentially every muon
implanted in such a sample sees a local hyperfine field.\cite{nach98}  There
appear to be no significant regions of the sample that do not show local
magnetic order.  An alternative model that allows local order everywhere
involves disorder in the spacing between charge stripes.  One assumes that the
charge stripes are separated by an integral number of lattice spacings, and
that at least two distinct stripe periods exist.  Such a model is consistent
with the idea that commensurability with the lattice should play a role in the
pinning of stripes in the LTT phase.  Furthermore, as we will see, it is
compatible with the incommensurate peak splitting,\cite{tran96b}
$\epsilon=0.118$.

Formulas for calculating the scattered intensity from a random mixture of 2
structural units were first worked out by Hendricks and
Teller.\cite{hend42,kaki52}  (One of us used these formulas
previously\cite{tran94b} to model staging disorder associated with interstitial
oxygens in La$_2$NiO$_{4+\delta}$.)   As structural units we take magnetic
unit cells of period $8a$ and $10a$.  (Each of these corresponds to two unit
cells of the corresponding charge modulation.)  Within each unit, we consider
only Cu sites, and assume, for simplicity, a sinusoidal variation of the spin
density.  The atomic displacements associated with the charge modulation are
also taken to be sinusoidal. 

\begin{figure}
\centerline{\psfig{figure=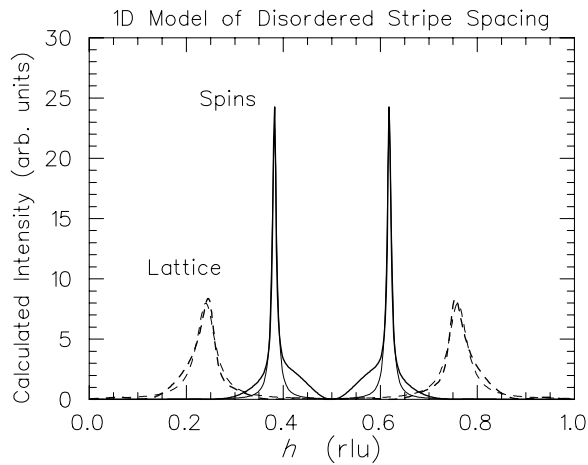,width=3.0in}\medskip}
\caption{Calculated scattered intensity as a function of $h$ for the model of
1D stripe-spacing disorder discussed in the text.  Thick solid line:
scattering due to spins; thick dashed line: signal due to lattice modulation. 
Thin lines represent fitted Lorentzians.}
\label{fg:disorder}
\end{figure}

Within the model, the only free parameter is the ratio of the occurrence
probabilities for the two structural units.  We adjusted this value to make
the peaks associated with spin modulation appear at positions consistent with
experiment.  This criterion was satisfied with a probability of 75\%\ for the
$8a$ unit and 25\%\ for the $10a$ unit.  The resulting scattered intensities
due to the spin and lattice modulations are shown in Fig.~\ref{fg:disorder} as
a function of momentum (in reciprocal lattice units) along the modulation
direction.  To extract peak positions and widths for comparison with
experiment, Lorentzians were fit to the calculated intensity.  The results are
listed in Table~\ref{tb:mod}.

It is obvious from the figure and the fitted values in the table that the
disorder assumed in the model yields finite peak widths that are larger for the
lattice peaks than for the spin peaks.  The width for magnetic peaks in this
simple model is about twice what we have found experimentally, and the same
discrepancy is found for the charge order peaks.\cite{vonz98}  However, the
ratio of calculated widths for lattice peaks relative to spin peaks ($=4$) is
exactly the same as the ratio of the experimentally determined
widths.\cite{vonz98}  The difference in widths for lattice and spin peaks is
easily understood: the scattering from the two distinct unit cells of the model
is more nearly in phase at the magnetic wave vector than at the charge-order
wave vector.  While one could tune the choice of unit cells in the model to
obtain better quantitative agreement with experiment, the salient point here is
that the experimentally observed peak widths and positions are qualitatively
consistent with disorder in the stripe spacing.

\begin{table}[tb]
\caption{Values for effective $\epsilon$ (from peak position) and half-width
at half maximum (HWHM) obtained from Lorentzian fits to the disordered-stripe
model calculation shown in Fig.~\protect\ref{fg:disorder}.}
\label{tb:mod}
\begin{tabular}{ccd}                                     
Peak & $\epsilon_{\rm eff}$ & HWHM \\  
     &   (rlu)     & ($10^{-3}$ \AA$^{-1}$) \\ 
\hline
Spin & 0.118 & 9.5 \\ 
Lattice & 0.120 & 38.3 \\
\end{tabular}
\end{table}

What causes the stripe disorder?  One possibility is the local potential
variations due to the randomly positioned divalent Sr ions.  On the other
hand, evidence has recently been presented\cite{attf98} that all variations in
cation radius on the La site contribute to the suppression of $T_c$, so
perhaps the substituted Nd, which is significantly smaller than the La, makes a
contribution to disorder comparable to that of the Sr.  The relevance of
disorder to pinning charge stripes and the development of glassy order is
discussed in Refs.~\onlinecite{kive98b,hass98}.

In applying the Hendricks-Teller model, we have considered disorder in only one
direction, perpendicular to the stripes.  If the disorder is induced by randomly
positioned Sr and Nd substituents, then there should also be defects, such as
dislocations, along the stripes.  Equal probabilities for defects in parallel
and perpendicular directions would explain the absence of a detectable
anisotropy in scattering peak width.

\subsection{Temperature-dependence of the correlation length}

We have seen that the magnetic peak width begins to grow for temperatures
above 30~K, the point at which $\mu$SR measurements\cite{nach98} indicate the
disappearance of static magnetic order.  Given the finite widths for both the
magnetic and charge-order peaks at low temperature, it is reasonable to ask
whether the decrease in the magnetic correlation might be due to increasing
disorder (perhaps dynamic) in the positions of the charge stripes.  According to
the Hendricks-Teller model applied above, such an increase in charge stripe
disorder should be reflected in a relative growth in the charge-order peak
width comparable to that of the magnetic peaks.  However, the charge-order
width observed in the X-ray diffraction measurements\cite{vonz98} shows little
variation up to 65~K.  Hence, it appears most likely that the charge stripe
configuration remains static throughout the region in which the
magnetic width is observed to vary.

It seems reasonable, then, to analyze the decrease of the magnetic
correlation length in terms of spin fluctuations within well defined
domains.  In section IV we considered the implications of an effective
anisotropic Heisenberg model first proposed by Castro Neto and
Hone.\cite{neto96,vand98}  The temperature dependence of $\kappa$ is well
described by the renormalized-classical behavior predicted by the analysis of
the model; however, the effective spin-wave velocity perpendicular to the
stripes implied by the fit to the model appears to be inconsistent with the
measurements of the spin fluctuations at $\hbar\omega=3$~meV.  The model can
be improved by considering the enhanced quantum effects that occur when the
magnetic domains correspond to even-legged spin ladders.  The latter model
appears to run into trouble when one tries to consider the doping dependence of
the magnetic correlations, since magnetic order is most
robust\cite{tran97a,kuma94,kimu98} at $x\approx\frac18$.

It is quite possible that the interaction between the holes in the domain walls
and the spins in the domains is not adequately characterized in terms of an
effective exchange coupling between antiferromagnetic domains.  An alternative
model of the interaction has been evaluated by Emery, Kivelson, and
Zachar.\cite{emer97}  They find that the interaction takes place by the
hopping of pairs of holes from a charge stripe into the magnetic regions, and
that the pair hopping tends to enhance singlet correlations.  This picture
would certainly be consistent with a weak effective coupling between
neighboring domains; however, a specific prediction for the low-energy spin
fluctuations relevant to the present case has not been made.

\subsection{Correlation between resistivity and structure}

For crystals of La$_{1.6-x}$Nd$_{0.4}$Sr$_x$CuO$_4$ with $x$ near $\frac18$,
the in-plane resistivity shows\cite{naka92} an abrupt upward jump on
transforming to the LTT structure near 70~K.  When Nd is replaced by Eu, the
LTT transition shifts up to approximately 130~K, and no jump in resistivity is
found.\cite{huck98}  Instead, on cooling there is a continuous deviation from
a linear temperature dependence, with a significant upturn below 50 to 60~K. 
In light of that, the correlation between resistivity and LTT-like volume
fraction above 80~K in our Nd-doped $x=0.15$ sample should perhaps come as no
surprise.

It seems reasonable to assume that the rise in resistivity on cooling is
associated with pinning of the charge stripes.  If the charge order were like
that in a conventional CDW system, in which a gap opens
over a significant portion of the Fermi surface, then one might expect to see
the charge order parameter directly reflected in the resistivity.  So far,
infrared reflectivity studies have provided no evidence for a charge
gap,\cite{taji98} and such a gap would appear to be incompatible with the
superconductivity observed in the $x=0.15$
sample.\cite{tran97a,oste97,mood97,nach98} On the other hand, the gradual
increase in resistivity is consistent with the glass-like stripe disorder that
was analyzed above.  Also, a recent study of the CDW system TaS$_2$ has shown
that a very small amount of disorder in that system can suppress the transition
to commensurate CDW order and the concomitant jump in resistivity.\cite{zwic98} 
It has been argued recently that transverse fluctuations of charge stripes are
important for suppressing CDW order along the stripes.\cite{kive98}  Perhaps
static disorder caused by randomly positioned dopants can also frustrate true
CDW order.

\acknowledgements

We are pleased to acknowledge stimulating discussions with R. J. Birgeneau, A.
H. Castro Neto, D. Hone, A. R. Moodenbaugh, G. Shirane, J. Zaanen, and
especially with V. J. Emery and S. A. Kivelson.  Assistance from J. D. Axe in
the early stages of this work is greatly appreciated. This research has received
some support from the U.S.-Japan Cooperative Neutron-Scattering Program.
Work at Brookhaven was carried out under
Contract No.\ DE-AC02-98CH10886, Division of Materials Sciences, U.S.
Department of Energy.



\end{document}